\documentclass[12pt]{article}

\usepackage{latexsym}
\usepackage{amsmath}
\usepackage{amssymb}
\usepackage{amsthm}
\usepackage{amscd}
\usepackage[mathscr]{eucal}
\usepackage{fullpage}
\usepackage{graphicx}
\usepackage{subfigure}
\usepackage{psfrag}
\usepackage{rotating}

\usepackage{color}

\newcommand{\textfrac}[2]{{\textstyle \frac{#1}{#2}}}

\newcommand{\R}{\mathbb{R}}

\newcommand{\A}{{\fontfamily{phv}\fontseries{b}\selectfont A}}
\newcommand{\B}{{\fontfamily{phv}\fontseries{b}\selectfont B}}
\newcommand{\C}{{\fontfamily{phv}\fontseries{b}\selectfont C}}
\newcommand{\D}{{\fontfamily{phv}\fontseries{b}\selectfont D}}

\definecolor{ggreen}{cmyk}{0.7,     0,      0.9,      0}
\definecolor{viol}{cmyk}{0.3,1,0,0}
\definecolor{myred}{cmyk}{0.1, 1, 0.5, 0}
\definecolor{bblue}{rgb}{0.2, 0.29996, 0.8 }

\theoremstyle{plain}

\begin{document}


\title{\bf Dynamics of Robertson-Walker\\ spacetimes with diffusion}

\author
{A.~Alho  \\
          {\small Centro de An\'alise Matem\'atica, Geometria e Sistemas Din\^{a}micos}  \\
     {\small Instituto Superior T\'ecnico, Lisboa, Portugal} \\     
      {\small  \tt   aalho@math.ist.utl.pt}\\[0.4cm]
      S. Calogero  \\
      {\small Department of Mathematical Sciences}  \\
      {\small Chalmers University of Technology, University of Gothenburg}  \\
      {\small Gothenburg, Sweden} \\
      {\small  \tt  calogero@chalmers.se}\\[0.4cm]
      M. P. Machado Ramos  \\
     {\small Departamento de Matem\' atica e Aplica\c{c}\~{o}es}  \\
     {\small Universidade do Minho, Guimar\~aes, Portugal}  \\
     {\small  \tt mpr@mct.uminho.pt}\\[0.4cm]
A. J. Soares  \\
     {\small Centro de Matem\' atica,}  \\
     {\small Universidade do Minho, Braga, Portugal} \\
     {\small  \tt   ajsoares@math.uminho.pt}   \\
}

\date{}
\maketitle


\begin{abstract}
\noindent
 
We study the dynamics of  spatially homogeneous and isotropic spacetimes containing a fluid undergoing 
microscopic velocity diffusion in a cosmological scalar field.   
After deriving a few exact solutions of the equations, we continue by analyzing the qualitative behavior of  
general solutions. To this purpose we recast the equations in the form of a two dimensional dynamical system 
and perform a global analysis of the flow. Among the admissible behaviors, we find solutions that are 
asymptotically de-Sitter both in the past and future time directions and which undergo accelerated expansion at 
all times.
%
%
%
%
%

\end{abstract}


\section{Introduction}
\label{sec:int}
In a recent paper~\cite{simone2} one of us (SC) introduced a new type of fluid matter model in general relativity, in which the fluid particles undergo microscopic velocity diffusion in a cosmological scalar field. 
The energy-momentum tensor $T^{\mu\nu}$ and the current density $J^\mu$ of the fluid are given by the usual expressions
\begin{equation}
T^{\mu\nu}=\rho u^\mu u^\nu +p(g^{\mu\nu}+u^\mu u^\nu ),  \qquad
J^\mu =n u^\mu,
\label{TJ}
\end{equation}
where $\rho$ is the rest-frame energy density, $p$ the pressure, $u$ the four-velocity and $n$ the particle density of the fluid. The diffusion behavior is imposed by postulating the equations 
\begin{subequations}
\begin{align}
&\nabla_\mu T^{\mu\nu}=\sigma J^\nu,\label{diffeq}\\
& \nabla_\mu(n u^\mu )=0,\label{consn}
\end{align}
\end{subequations}
where $\sigma>0$ is the diffusion constant, which measures the energy gained by the particles  per unit of time due to the action of the diffusion forces. Equation~\eqref{consn} expresses the usual conservation of the total number of fluid particles, while~\eqref{diffeq} is the diffusion equation. In fact it was shown in~\cite{simone1,simone2} that~\eqref{diffeq} is the (formal) macroscopic limit of a Fokker-Planck equation on the kinetic particle density, which is a standard model for diffusion dynamics taking place at the microscopic level on the particles velocity~\cite{ri}.

By projecting~\eqref{diffeq} along the direction of $u^\mu$ and onto the hypersurface orthogonal to $u^\mu$, we obtain the following equations on the matter fields:
\begin{subequations}\label{mattereq}
\begin{eqnarray}
& \nabla_{\mu} (\rho u^{\mu})+p\nabla_{\mu} u^{\mu}=\sigma n,\label{ProjTJ1}\\
&(\rho +p) u^\mu \nabla_{\mu}u^\nu +u^\nu u^\mu \nabla_{\mu}p+\nabla^{\nu} p=0,\label{ProjTJ2}\\
&\nabla_\mu(n u^\mu )=0.\label{ProjTJ3}
\end{eqnarray}
\end{subequations}
The system~\eqref{mattereq} on the matter variables may be completed by assigning an equation of state between the pressure and the energy density (barotropic fluid). In this paper we assume that
\begin{equation}
p=(\gamma -1 )\rho,
\label{prholinear}
\end{equation}
for some constant $2/3<\gamma<2$, so that in particular the fluid satisfies the strong and dominant energy conditions. 
The case $\gamma=1$ gives rise to a dust fluid, while $\gamma=4/3$ corresponds to a radiation fluid.

Since the energy-momentum is not divergence-free, the coupling of the Einstein equations with the diffusion equation~\eqref{diffeq} is incompatible with the (contracted) Bianchi identity 
\begin{equation}\label{bianchi}
\nabla^\mu \left(R_{\mu\nu}-\frac{1}{2}g_{\mu\nu}R\right)=0.
\end{equation}
This incompatibility can be resolved by postulating the existence of an additional matter field interacting with the fluid particles and restoring the local conservation of energy. This new matter field plays the role of the background medium in which the particles undergo diffusion. The simplest model for this medium is a vacuum-energy source described by a cosmological scalar field, which leads to the following Einstein equations (in units $8\pi G=c=1$):
\begin{equation}
R_{\mu\nu}-\frac{1}{2}g_{\mu\nu}R+\phi g_{\mu\nu}=T_{\mu\nu}.
\label{einsteinequations}
\end{equation}
The evolution equation on the scalar field $\phi$ determined by~\eqref{einsteinequations}, the Bianchi identity~\eqref{bianchi}, and the diffusion equation~\eqref{diffeq} is
\begin{equation}\label{phieq}
\nabla_\nu \phi =\sigma J_\nu.
\end{equation}
Notice that when $\sigma=0$, the model under study reduces to the Einstein-perfect fluid system with cosmological constant. The latter has important applications in cosmology, where, under appropriate symmetry assumptions, leads to the $\Lambda$CDM model, currently the most popular cosmological model of the universe. The analogous cosmological model with diffusion is the $\phi$CDM model, in which $\Lambda$ is replaced by the ``variable cosmological constant" $\phi$, representing a dark energy field interacting with dark matter by diffusion. This cosmological model has been studied in detail in~\cite{CV}, where in particular an upper bound on the diffusion constant $\sigma$ has been estimated in order for the model to be compatible with the current cosmological observations. We emphasize that the interaction between the dark energy field and the matter fluid in the diffusion model is a natural and compelling consequence of the relativistic diffusion mechanism. Therefore the diffusion model  
considered here is fundamentally different from the other interacting models existing in the literature, where the interaction between dark energy and dark matter is supported only on a pure phenomenological basis; see~\cite{CV} for a more detailed discussion on this important point. Further studies on the diffusion model described above  can be found in \cite{shogin, shogin2, shogin3}.   

In the present work we derive a complete characterization of the diffusion dynamics  
in a spatially homogeneous and isotropic spacetime.
Using standard Robertson-Walker coordinates $(t,r,\theta,\varphi)$, we write the metric as
\begin{equation}\label{RW}
ds^2=-dt^2+a(t)^2\left[\frac{dr^2}{1-k r^2}+r^2d\Omega^2\right],\quad d\Omega^2=d\theta^2+\sin^2\theta d\varphi^2,
\end{equation}
where $k=0,\pm 1$ is the curvature parameter.
%
%
The equations for the scale factor $a(t)$, the energy density $\rho(t)$ of the fluid and
the cosmological scalar field $\phi(t)$ are
\begin{subequations}\label{system}
\begin{align}
&\dot{a}=Ha,
   \label{hubble}\\
&\dot{\rho}=-3\gamma H\rho-\dot{\phi},
   \label{rhodot}\\
&\dot{\phi}=-\sigma n_0 \left(\frac{a_0}{a(t)}\right)^{3},
   \label{phidot}\\
&\dot{H}=\frac{1}{3}\left[\phi-\left(\frac{3}{2}\gamma-1\right)\rho\right]-H^2,
\label{Hdot}\\
&H^2=\frac{1}{3}(\rho+\phi)-\frac{k}{a(t)^2},\label{constraint}
\end{align}
\end{subequations}
where $a_0>0$, $n_0>0$ are the values of the scale factor and the particle density at the time $t=0$. 
The behavior of solutions to the preceding system for $k=0$ has been investigated numerically in~\cite{simone2}. 
In the present paper we provide analytical proofs for these numerical results and many  others 
which were not discovered in~\cite{simone2}. For instance we show the 
existence of solutions to the Einstein equations~\eqref{system} which exhibit accelerated expansion for 
all times, thereby providing a unified cosmological model for primordial inflation and late-time acceleration  of the universe. 
Other interesting new types of solutions found in this paper  include solutions for $k=-1$ which can be matched to anti-de-Sitter 
at a finite time in the past and solutions for $k=1$ which describe an initially contracting, then expanding and finally recollapsing spacetime. 
These are only examples of the new and rich dynamics of the diffusion model discovered in the present paper.

The proof of our results is based on the application of techniques from the theory of finite dimensional dynamical 
systems, which is by now a standard tool in theoretical cosmology.  
An extensive discussion on this method for perfect fluid cosmological models without diffusion ($\sigma=0$) and 
with $\Lambda=0$ can be found in \cite{WE}. The dynamical systems approach has also been applied for the analysis 
of cosmological models with positive cosmological constant \cite{GE99} and with matter sources other than 
fluids~\cite{CH, HU}.
When the cosmological constant is replaced by a minimally coupled scalar field, the case with exponential 
potential has been studied in~\cite{Coley}. Only recently an efficient dynamical systems approach has been 
introduced for monomial potentials~\cite{AU2015,AHU2015}. 
See also \cite{F98a,F98b,U13}, and references 
therein, for other types of potential. 

The rest of the paper is organized as follows. In Section~\ref{sec:eqs} we solve the system~\eqref{system} 
explicitly for the case where the scale factor $a(t)$ in the metric is linear on time. Moreover we point out 
the possible existence of solutions which can be matched to vacuum solutions in the past of some time $t_0$ and 
introduce a classification of solutions to~\eqref{system} based on their asymptotic behavior toward the future 
and past time directions. The fact that solutions with such behaviors do exist is proved rigorously in 
Section~\ref{sec:ds}. The proof is obtained by rewriting~\eqref{system} using new dynamical variables, 
in such a way that the admissible self-similar asymptotic states of the system are represented by fixed points 
of the dynamical system in the new variables. The dynamical system in question is two dimensional but, as opposed 
to the diffusion-free case, it is not defined on a compact state-space. We remark that the new variables employed 
here are not obtained by normalizing the old ones with the Hubble function, as those used for instance in~\cite{shogin, shogin2, shogin3}, 
and therefore we do not restrain our models to be forever expanding or contracting. This explains the considerably 
richer dynamics found in our investigation compared to the earlier available results on this model.
%

\section{Classification of cosmological models}
\label{sec:eqs}
The initial data set for the system~\eqref{system} consists of $(a_0,\rho_0,H_0,\phi_0)$, where $a_0$, $H_0$, $\rho_0$ and $\phi_0$ satisfy~\eqref{constraint} at time $t=0$, i.e.,
\begin{equation}
\label{initialconstraint}
H_0^2=\frac{1}{3}(\rho_0+\phi_0)-\frac{k}{a_0^2}.
\end{equation}
We remark that we do not necessarily identify $t=0$ with the present time. For $k=0,\pm 1$, we denote by $\mathcal{I}_k$ the three-dimensional manifold of admissible initial data, i.e.,
\[
\mathcal{I}_k=\{(a_0,\rho_0,H_0,\phi_0)\in (0,\infty)\times (0,\infty)\times\R\times\R: \text{\eqref{initialconstraint} holds}\} .
\] 
Moreover we denote by $q$ the deceleration parameter:
\begin{equation}\label{q}
q=-\frac{a \ddot a}{\dot{a}^2}=-1-\frac{\dot{H}}{H^2}.
\end{equation}
We shall say that a solution of the Einstein equations~\eqref{system} is expanding with acceleration if $H>0$ and $q<0$, and that it is contracting with acceleration if $H<0$ and $q<0$. 
Moreover, we say that a solution is singular at some time $t_0$ if $a(t_0)=0$; we do not discuss here the question of whether the corresponding spacetime singularity is a curvature 
singularity or just a coordinate singularity.

The main purpose of this section is to introduce a classification of the solutions to the Einstein equations~\eqref{system}  
based primarily on their asymptotic behavior toward the past and the future. We first introduce a number of exact solutions.
\subsection{Exact solutions}
Given the large variety of numerical solutions discovered in~\cite{simone2}, it is unlikely that one can 
solve explicitly the system~\eqref{system} for general initial data.  
However, in the particular case that the scale factor $a(t)$ is linear on time,
we find the following explicit solution:
\begin{subequations}
\label{solution}
\begin{align}
& a(t)=a_0+\delta_k t,\\
& \phi(t)=\frac{3\beta}{2\delta_k} a(t)^{-2},\\
&  \rho(t)=\frac{3\beta}{\delta_k (3\gamma-2)}a(t)^{-2},
\end{align}
\end{subequations}
where we set 
\[
\beta=\frac{\sigma n_0 a_0^3}{3}
\]
and $\delta_k$ is the real solution of the polynomial equation
\begin{equation}\label{eqalpha}
\delta^3+  k\delta-\frac{3\beta\gamma }{2(3\gamma-2)}=0.
\end{equation}
Note that $\delta_k>0$, for all $k=0,\pm1$.
{\color{black} In particular,}
for $k=0$ the solution {\color{black}(\ref{solution})} becomes
\begin{subequations}\label{solution2}
\begin{align}
& a(t)=a_0+\left(\frac{3\beta\gamma}{2(3\gamma-2)}\right)^{1/3}t,\\
& \phi(t)=\left(\sqrt{\frac{3\gamma-2}{\gamma}}\frac{3\beta}{2}\right)^{2/3}a(t)^{-2},\\
& \rho(t)=\left(\frac{2}{\gamma}\right)^{1/3}\left(\frac{3\beta }{3\gamma-2}\right)a(t)^{-2}.
\end{align}
\end{subequations}
For $\sigma=0$ the solution~\eqref{solution2} reduces to the Minkowski spacetime, while the solution~\eqref{solution} for $k=-1$ reduces to the Milne spacetime. 
In contrast, there is no diffusion-free analogue of the solution~\eqref{solution} for $k=1$. 

All solutions defined by~\eqref{solution} are singularity free and forever expanding in the future, while in the past they become singular at the time $t_-=-a_0/\delta_k$. 
The future expansion takes place at a constant rate. In particular, the deceleration parameter defined by~\eqref{q} is zero for the solutions~\eqref{solution}. 
We also remark that the singularity at $t=t_-$ is a curvature singularity. In fact, computing the Ricci scalar of the solution~\eqref{solution} we obtain, using~\eqref{einsteinequations}, 
\[
R=4\phi+(4-3\gamma)\rho=\frac{9\beta\gamma}{\delta_k(3\gamma-2)}a(t)^{-2}\to +\infty,\quad \text{as }\ a(t)\to 0^+.
\]

\subsection{Vacuum solutions}\label{vacsec}

Next we discuss the vacuum solutions, i.e., solutions for which $\rho(t)=n(t)=0$ for all times. Such solutions are important because they may act as future/past attractors of general solutions of the system~\eqref{system}. Since $\dot\phi=0$ when $\rho=n=0$, vacuum solutions correspond to maximally symmetric vacuum spacetimes with cosmological constant $\phi=\Lambda$. It is well known that any such spacetime is either de-Sitter (if $\Lambda>0$), anti-de-Sitter (if $\Lambda<0$), or Minkowski (if $\Lambda=0$).  We also recall that the coordinates $(t,r,\theta,\varphi)$ cover the full de-Sitter space for $k=1$, while in all other cases they cover only a portion of (anti)-de-Sitter space. For more background on the geometry of de-Sitter/anti-de-Sitter we refer to~\cite{Gib, Kim}.  

Letting $\rho=n=0$ and $\phi=\Lambda$ in the system~\eqref{system}, we obtain that the scale factor of vacuum solutions satisfies
\begin{equation}\label{vacuumeq}
\frac{\ddot a}{a}-\left(\frac{\dot a}{a}\right)^2=\frac{k}{a^2},\quad a(0)=a_0,\quad \dot{a}(0)=H_0 a_0,\quad H_0=\pm\sqrt{\frac{\Lambda}{3}-\frac{k}{a_0^2}}.
\end{equation}
The explicit form of the solution to~\eqref{vacuumeq} depends on the values of the parameters $k$ and $\Lambda$. Although the form of the scale factor in all various cases is well-known, we give it below for easy reference. 

For $k=0$ we must have $\Lambda\geq 0$ and the scale factor is given by
\begin{equation}\label{flatdesitter}
a(t)=a_0e^{H_0 t},\quad\text{where either $H_0=\sqrt{\frac\Lambda3}$,\quad or\quad $H_0=-\sqrt{\frac\Lambda3}$}.
\end{equation}
The corresponding spacetime is de-Sitter.  
For $\Lambda=0$ (i.e., $H_0=0$) both solutions reduce to $a(t)=a_0$, and the corresponding spacetime is Minkowski. 

For $k=1$  we must have $\Lambda\geq 3/a_0^2$. The solutions  are given by
\begin{subequations}\label{closeddeSitter}
\begin{align}
&a(t)=\sqrt{\frac{3}{\Lambda}}\,\mathrm{cosh}\Bigg[\sqrt{\frac{\Lambda}{3}}\,t+\log\Big[a_0\Big(\sqrt{\frac{\Lambda}{3}}+H_0\Big)\Big]\Bigg],\\[0.5cm]
&\text{where either $H_0=\sqrt{\frac\Lambda3-\frac{1}{a_0^2}}$,\quad or\quad  $H_0=-\sqrt{\frac\Lambda3-\frac{1}{a_0^2}}$}.
\end{align}
\end{subequations} 
The two solutions coincide when $\Lambda=3/a_0^2$ (i.e., $H_0=0$) and become
\begin{equation}\label{closeddeSitterspecial}
a(t)=a_0\,\mathrm{cosh}\Big(\sqrt{\frac\Lambda3\,t}\Big).
\end{equation}
The corresponding spacetime is de-Sitter.

For $k=-1$, the solutions are different according to the sign of $\Lambda$. For $\Lambda> 0$ we obtain 
\begin{subequations}\label{opendeSitter}
\begin{align}
&a(t)=\sqrt{\frac{3}{\Lambda}}\,\mathrm{sinh}\Bigg[\sqrt{\frac{\Lambda}{3}}\,t+\log\Big[a_0\Big(\sqrt{\frac{\Lambda}{3}}+H_0\Big)\Big]\Bigg],\\[0.5cm]
&\text{where either $H_0=\sqrt{\frac\Lambda3+\frac{1}{a_0^2}}$,\quad  or\quad  $H_0=-\sqrt{\frac\Lambda3+\frac{1}{a_0^2}}$}.
\end{align}
\end{subequations} 
The corresponding spacetime is again de-Sitter.

For $k=-1$ and $\Lambda=0$ we obtain the Milne solutions:
\begin{equation}\label{milne}
a(t)=a_0(1+H_0t),\quad\text{where either $H_0=\frac{1}{a_0}$,\quad or\quad $H_0=-\frac{1}{a_0}$.}
\end{equation} 
For $-3/a_0^2\leq \Lambda<0$, the solutions to~\eqref{vacuumeq} are given by
\begin{subequations}\label{antidesitter}
\begin{align}
& a(t)=\sqrt{\frac{3}{|\Lambda|}}\sin\Bigg[\mathrm{arccot}\Big(\sqrt{\frac{3}{|\Lambda|}}H_0\Big)+\sqrt{\frac{|\Lambda|}{3}t}\Bigg],\\
&\text{where $H_0=\pm\sqrt{\frac\Lambda3+\frac{1}{a_0^2}}$},
\end{align}
\end{subequations}
and for $H_0=0$ they both reduce to
\begin{equation}
a(t)=\sqrt{\frac{3}{|\Lambda|}}\cos\Big(\sqrt{\frac{|\Lambda|}{3}t}\Big).
\label{sincos}
\end{equation} 
The corresponding spacetime is anti-de-Sitter.

\subsection{Vacuum matching solutions}\label{vacmatsec}
A remarkable difference with the diffusion-free case is the possibility that the energy density $\rho$ vanished at some finite time $t_0$ while the scale factor $a$ is still regular. This possibility arises because $\rho=0$ is not a solution of~\eqref{rhodot}. By a time translation we may set $t_0=0$ for any given such type of solution and therefore we assume that
\[
\rho(0)=0,\quad a(0)>0, \quad |H(0)|<\infty.
\]
By~\eqref{rhodot}, $\dot{\rho}(0)>0$, hence $\rho(0)<0$ for $t<0$. To avoid this unphysical region of spacetime, 
we replace it with either the de-Sitter or the anti-de-Sitter, depending on the sign of $\phi(0)$. This extended 
spacetime is singularity free in the past and in the region $t\leq 0$ covered by the coordinates $(t,r,\theta,\phi)$, the scale factor is given by one of the expressions~\eqref{flatdesitter},~\eqref{closeddeSitter},~\eqref{opendeSitter}, depending on the value of $k$, $\phi(0)=\Lambda$ and $H_0=H(0)$. 
The matching at $t=0$ is $C^2$ in the scale factor $a(t)$ and $C^0$ in the scalar field $\phi(t)$ and in the energy density $\rho(t)$.

\subsection{Classification of general solutions}\label{genclas}
To conclude this section we introduce a classification of general solutions to~\eqref{system} based on their asymptotic behavior toward the past and future time directions. In particular, we say that a solution is of type \A\  if it becomes singular at some finite time in the past, while in the future it is singularity free and asymptotically de-Sitter. A solution is said to be of type \B\ if it can be matched to a de-Sitter solution at some finite time in the past, while in the future it is singularity free  and asymptotically de-Sitter. Solutions of type \C\ are those which can be matched to a vacuum solution at some finite time in the past and which become singular at some finite time in the future. Finally, a solution of type \D\ is a solution that becomes singular at finite time in both time directions. 

The main result of this paper can be formally stated as follows: For all $k=0,\pm 1$, there exists four disjoint three-dimensional submanifolds of initial data $\mathcal{A}_k\subset \mathcal{I}_k$, $\mathcal{B}_k\subset\mathcal{I}_k$, $\mathcal{C}_k\subset\mathcal{I}_k$, $\mathcal{D}_k\subset\mathcal{I}_k$ such that if the initial data belong to $\mathcal{A}_k$, the corresponding solution of~\eqref{system} is of type \A, if the initial data belong to $\mathcal{B}_k$, the corresponding solution is of type \B, etc. Moreover solutions which are not launched by initial data in the set $\mathcal{A}_k\cup\mathcal{B}_k\cup\mathcal{C}_k\cup\mathcal{D}_k$ are atypical, that is to say, they correspond to initial data forming a two-dimensional submanifold  of $\mathcal{I}_k$. 

\section{Qualitative dynamics of solutions}
\label{sec:ds}
For an introduction to dynamical systems theory see~\cite{Perko}.
Let
\begin{equation}
D=\sqrt{H^2+\frac{1}{a^2}}
\label{defD}
\end{equation}
and define a new time variable $\tau$  by
\[
\frac{d}{d\tau} \, (\cdot) = \frac{1}{D} \frac{d}{dt} \, (\cdot); 
\]
in the following we use the notation $(\cdot )' = \frac{d}{d\tau}(\cdot)$. We also introduce the dimensionless variables
\begin{equation}
H_{D}      =\frac{H}{D}, \quad 
M_{D}      =\frac{1}{aD}, \quad
\Omega_{D} = \frac{\rho}{3D^2}, \quad
Y_{D}      = \frac{\phi}{3D^2}, \quad
X_{D}      = \frac{\dot{\phi}}{3D^3}.
\label{chis1}
\end{equation}
These variables satisfy the algebraic constraints 
\begin{equation}\label{constnew}
 H^{2}_{D}+M^{2}_{D}=1,\quad X_D+\beta M^{3}_{D}=0,
\quad H^{2}_{D}=\Omega_{D}+Y_{D}-k M^{2}_{D},  
 \end{equation}
where we recall that $\beta=\sigma n_0 a^{3}_{0}/3$.
The first equation in~\eqref{constnew} follows by the definition of the new variables, 
while the second and third equations correspond 
respectively to~\eqref{phidot} and~\eqref{constraint}.
  
The evolution equation for the dimensional variable $D$ is given by
\begin{equation}\label{Deq}
 D^{\prime}=-H_{D}D (1+qH_D^2),
\end{equation}
while the remaining variables satisfy 
%


\begin{align*}
H^{\prime}_{D} & = -qH_D^2(1-H_D^2), \\
M^{\prime}_{D} & = qH_D^3M_D,  \\
\Omega^{\prime}_{D} &= -X_{D}+2H_D\Omega_{D}(1+qH_D^2-\textfrac{3}{2}\gamma), \\
Y^{\prime}_{D} & = X_{D}+2H_{D}Y_D(1+qH_D^2),\\
X^{\prime}_{D} & = 3qH_D^3X_{D},
\end{align*}
where 
\begin{equation}\label{qnew}
qH_D^2=-Y_D+\Big(\frac{3}{2}\gamma-1\Big)\Omega_D.
\end{equation}

Note that the equation on $D$ decouples from the rest of the system; this is due to the fact that $D$ is dimensional, while the other variables have no physical dimension (in our units). 

From the constraint equations \eqref{constnew}, the dimension of the dynamical system can be reduced by three, obtaining a
$2$-dimensional dynamical system. We choose to work with the variables $(Y_{D},H_{D})$. 
Once these are known, the  remaining variables can be determined globally from the constraints.  
The reduced dynamical system we shall be working with is then given by
\begin{subequations}\label{redyns}
\begin{align}
H^{\prime}_{D} & = \big[Y_{D}+(1-\textfrac{3}{2}\gamma)\Omega_{D}\big](1-H^{2}_{D}),\label{HDeq} \\
Y^{\prime}_{D} & = -\beta (1-H^{2}_{D})^{\textfrac{3}{2}}+2H_{D}Y_{D}\big[1-Y_{D}-(1-\textfrac{3}{2}\gamma)\Omega_{D}\big],
\end{align}
where
\begin{equation}\label{omega}
 \Omega_{D}=H^{2}_{D}+k(1-H^{2}_{D})-Y_{D}.
\end{equation}
\end{subequations}

As opposed to the standard diffusion-free case (which is obtained by setting $\beta=0$ in the equations above), 
the variable $\Omega_D$ is not bounded and $\Omega_D=0$ is not an invariant boundary. However,
since 
\[
(\Omega'_D)_{|_{\Omega_D=0}}=-X_D=\beta (1-H_D^2)^{3/2}>0,
\]
the curve $\Omega_D=0$ acts as ``semipermeable membrane":   the 
flow can cross this line only in one direction. In particular, the region $\Omega_D>0$ is future invariant 
and if an orbit in the region $\Omega_D>0$ intersects the vacuum line $\Omega_D=0$ in the past, then  $\Omega_D$ is negative for all earlier times along this orbit. As discussed in Section~\ref{vacmatsec}, the solutions of the original system~\eqref{system} corresponding to these orbits can be matched to a suitable vacuum solution at the time when the boundary $\Omega_D$ is crossed. We emphasize once again that this matching is not possible in the diffusion-free case, because $\Omega_D=0$ is an invariant boundary when $\beta=0$. 

By the preceding remarks, we only need to worry about the region $\Omega_D>0$, which in terms of the variables $(Y_D,H_D)$ means  
\begin{equation}\label{positiveOmega}
\Omega_D>0 \Leftrightarrow Y_D<H_D^2+k(1-H_D^2).
\end{equation}

The state-space $\mathcal{X}$ for the reduced dynamical system~\eqref{redyns} is then
\[
\mathcal{X}=\left\{(Y_{D},H_{D})\in\mathbb{R}\times\left(-1,1\right)\;:\; \Omega_{D}>0\right\}.
\]
Moreover the dynamical system admits a continuous extension on the boundary $H_D=\pm 1$.

Our next goal is to study the qualitative behavior of the flow of the dynamical system~\eqref{redyns} and to present the physical interpretation of this analysis in terms of solutions of the 
Einstein equations~\eqref{system} and their asymptotic behavior.

\subsection{Fixed points}
The dynamical system~\eqref{redyns} possesses five fixed points,
four of which are located on the boundary $(H_{D}=\pm1)$ and one in the interior, see Table~\ref{fixedpointsrooftable}.
The interior fixed point $\mathrm{S}_{(k)}$ is associated to the self-similar solutions~\eqref{solution2}
{\color{black} which have been characterized in Section \ref{sec:eqs},
with $a(t)$ being a linear function on time.}
Since the other fixed points are located at the boundary of the state-space,
they no longer correspond to exact solutions of~\eqref{system},
but to limiting states when one or more variables take an extreme value.

The main properties of 
the fixed points of the dynamical system~\eqref{redyns} are the following.
\begin{itemize}
\item[$\mathrm{S}_{(k)}$:]
At this fixed point we have 
\[
Y_{D}=\Big(1-\frac{2}{3\gamma}\Big)\Big(k+(1-k)\frac{\delta_k^2}{1+\delta_k^2}\Big)
\]
and 
\[
H_{D}=\frac{\delta_k}{\sqrt{1+\delta_k^2}},
\]
where $\delta_k$ is the (real, positive) solution of~\eqref{eqalpha} depending on $k$. Hence this fixed point identifies the self-similar solutions~\eqref{solution}. The fixed point $\mathrm{S}_{(k)}$  is hyperbolic and a simple local stability analysis shows that  $\mathrm{S}_{(k)}$ is a saddle fixed point.


\item[$\mathrm{F}_\pm$:]
At these fixed points we have $Y_{D}=0$ and $H_{D}=\pm1$. 
The fixed point $\mathrm{F}_+$ is a source of interior orbits. To see this, consider the flow induced by~\eqref{redyns} on the boundary $H_D=1$. This flow is described by the one-dimensional dynamical system
\[
Y'_D=3\gamma Y_D(1-Y_D).
\]
As $Y'_D>0$ for $Y_D\in (0,1)$ and $Y'_D<0$ for $Y_D<0$, the fixed point $\mathrm{F}_+$ repels the orbits on the boundary $H_D=1$. Moreover, using that 
\[
\left(\frac{d}{d\tau}\log(1-H_D^2)\right)_{(Y_D,H_D)=(0,1)}=3\gamma-2>0,
\]
we obtain that the variable $H_D$ is decreasing along interior orbits approching  $\mathrm{F}_+$. It follows that the fixed point  $\mathrm{F}_+$  repels interior orbits as well and thus that it is a source of interior orbits. We remark that since $\mathrm{F}_+$  lies on the boundary of the state space, this result cannot be obtained by linearizing the system around this fixed point. A similar argument shows that  $\mathrm{F}_-$ is a sink of interior orbits.
From the constraints~\eqref{constnew} we obtain $M_D=X_D=0$ and $\Omega_D=1$. 
Substituting in~\eqref{Deq} and going back to the original time variable $t$ we obtain
\begin{equation}\label{ddotas}
\dot {D}=\mp\frac{3}{2}\gamma D^2.
\end{equation}
It follows that along the solutions of the system~\eqref{system} which correspond to the interior orbits converging to (resp. emanating from) the fixed point $\mathrm{F}_-$ (resp. $\mathrm{F}_+$), the variable $D(t)$ blows-up in finite time toward the future (resp. past). For any such solution we may choose the origin of time so that the singularity occurs at $t=0$ and integrating~\eqref{ddotas} we find

\[
D(t)\sim \pm\frac{\frac{2}{3\gamma}}{t},\quad\text{as $t\to 0^\pm$.}
\] 
Since $H_D\to\pm 1$ implies $H(t)\sim \pm D(t)$, we obtain
\begin{equation}
H(t)\sim \frac{\frac{2}{3\gamma}}{t},\quad\text{as $t\to 0^\pm$.}
\label{H31}
\end{equation}
Thus orbits that emanate from the fixed point $\mathrm{F}_+$ (resp. converge to $\mathrm{F}_-$) identify solutions of the Einstein equations which behave like the expanding (resp. contracting) diffusion-free perfect fluid solution
with zero cosmological constant (i.e., the Friedmann-Lema\^{i}tre solution) in the limit toward the past (resp. future) singularity. 



\item[$\mathrm{dS}_\pm$:]
At these fixed points we have $Y_{D}=1$
and $H_{D}=\pm 1$, hence~\eqref{constnew} gives $M_D=X_D=\Omega_D=0$.     
An argument similar to the one used for the fixed point $\mathrm{F}_+$  shows that the fixed point $\mathrm{dS}_-$ is a source, while $\mathrm{dS}_+$ is an attractor of interior orbits.
By~\eqref{Deq} we obtain $D(t)= c$, which implies that $H(t)=\pm c$, where $c$ is a positive constant. It follows that the orbits that converge to the fixed point $\mathrm{dS}_+$ (resp. emanate from $\mathrm{dS}_-$) correspond to solutions of the Einstein equations which are singularity free in the future (resp. past) and behave like the expanding (resp. contracting) de-Sitter solution for late (resp. early) times for $t\to +\infty$ (resp. $t\to -\infty$).
\end{itemize}

\begin{table}[ht!]
\begin{center}
\begin{tabular}{|c|c|c|c|}
\hline  & & \\[-2ex]
{\bf Fixed point} & {\mathversion{bold}$Y_{D}$} & {\mathversion{bold}$H_{D}$} \\ [1ex]
\hline\hline & & \\[-2ex]
$\mathrm{dS}_{-} $ & $1$ & $-1$\\[1ex]
$\mathrm{dS}_{+} $ & $1$ & $\,\,\,\,1$\\[1ex]
$\mathrm{F}_{-}  $ & $0$ & $-1$\\[1ex]
$\mathrm{F}_+    $ & $0$ & $\,\,\,\,1$\\[1ex]
$\mathrm{S}_{k}  $ & $ (1-\textfrac{2}{3\gamma})(k+(1-k)\textfrac{\delta_k^2}{1+\delta_k^2})$ & $\textfrac{\delta_k}{\sqrt{1+\delta_k^2}}$ \\[1ex]
\hline
\end{tabular}
\end{center}
\caption{Fixed {\color{black} points} of the dynamical system~\eqref{redyns} in the state-space $\overline{\mathcal{X}}$. $\delta_k$ is the solution of~\eqref{eqalpha}.}
\label{fixedpointsrooftable}
\end{table}

\subsection{Analysis of the flow}
\begin{figure}[Ht!]
\begin{center}
\psfrag{A}[cc][cc][0.8][0]{\A}
\psfrag{B}[cc][cc][0.8][0]{\B}
\psfrag{C}[ll][cc][0.8][0]{\C}
\psfrag{D}[cc][cc][0.8][0]{\D}
\psfrag{H}[bc][cc][1][0]{$H_D$}
\psfrag{Y}[rr][bb][1][0]{$Y_D$}
\psfrag{F+}[cc][cc][0.8][0]{$\mathbf{\mathrm{F}_+}$}
\psfrag{F-}[cc][cc][0.8][0]{$\mathbf{\mathrm{F}_-}$}
\psfrag{dS+}[cc][cc][0.8][0]{$\mathbf{\mathrm{dS}_+}$}
\psfrag{dS-}[cc][cc][0.8][0]{$\mathbf{\mathrm{dS}_-}$}
\psfrag{S}[cc][cc][0.8][0]{$\mathbf{\mathrm{S}_{(0)}}$}
\subfigure[$k=0$]{\includegraphics[width=0.45\textwidth]{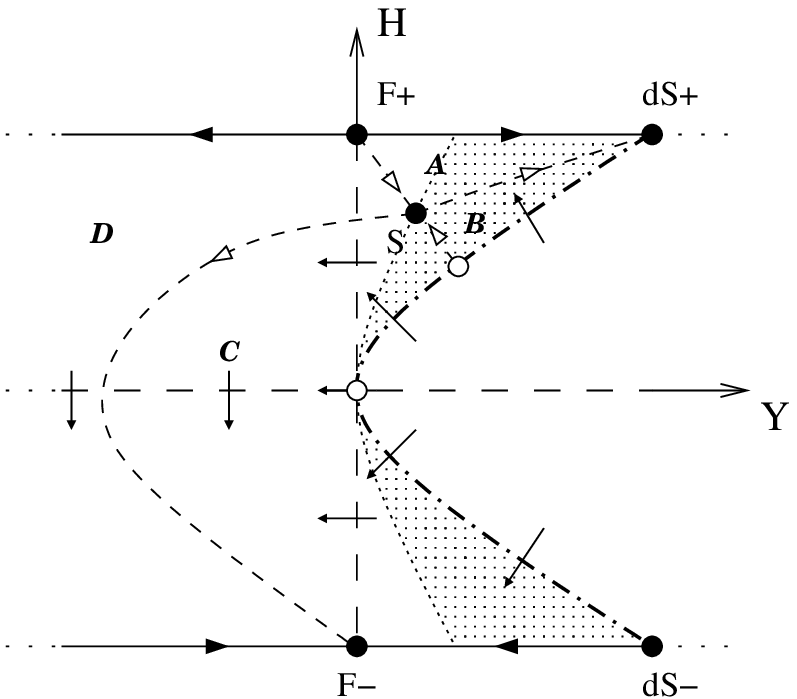}}\quad
\psfrag{S-}[cc][ll][0.8][0]{\quad$\mathbf{\mathrm{S}_{(\!-\!1)}}$}
\psfrag{$S_{(+1)}$}[rr][cr][0.8][0]{\quad$\mathbf{\mathrm{S}_{(\!+\!1)}}$}
\psfrag{A}[lc][cc][0.8][0]{\A}
\psfrag{B}[cc][cc][0.8][0]{\B}
\subfigure[$k=1$]{\includegraphics[width=0.45\textwidth]{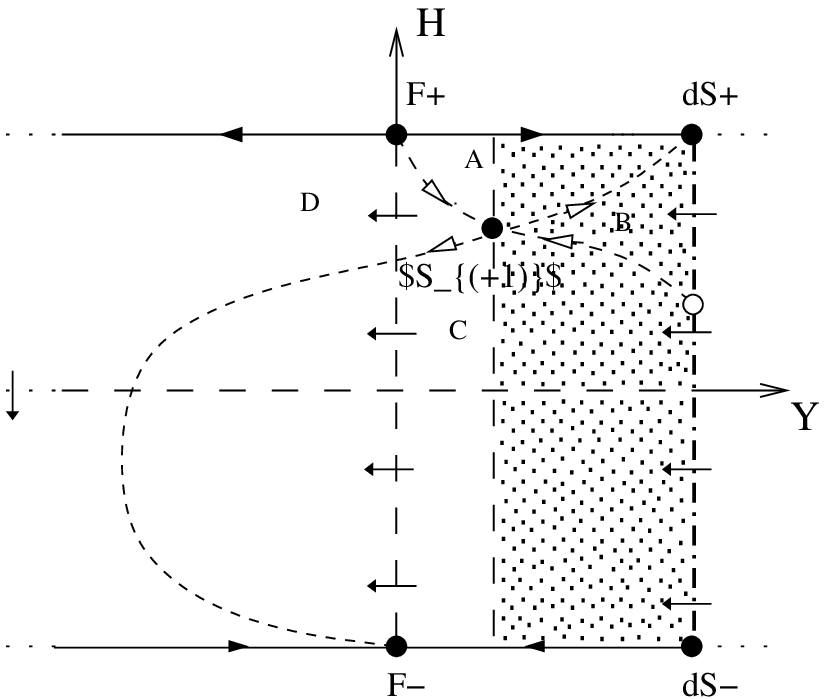}}
\subfigure[$k=-1$]{\includegraphics[width=0.7\textwidth]{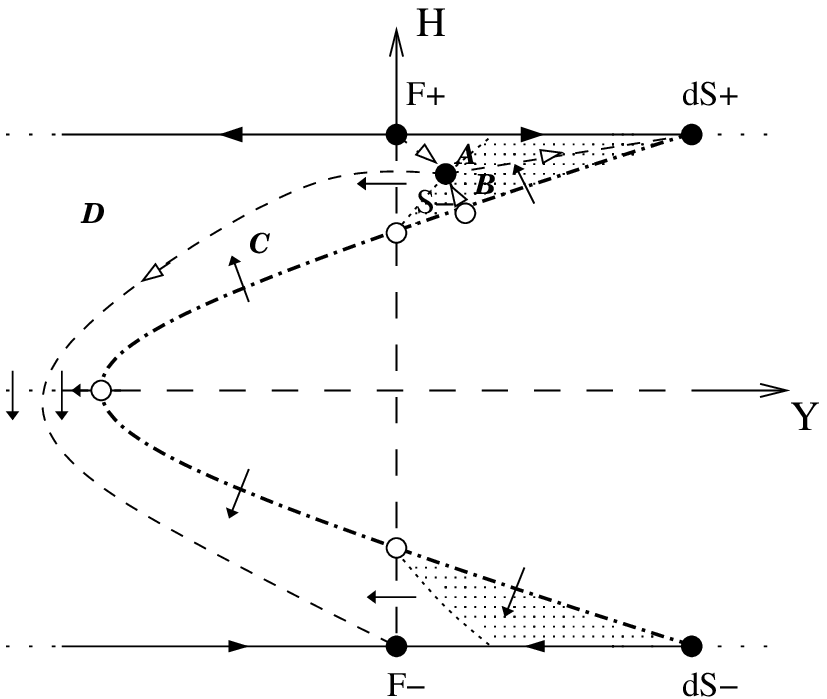}}\\
\end{center}
\caption{Partition of the state-space}\label{fig1}
\end{figure}

We begin by pointing out a few general properties of the flow of the dynamical system~\eqref{redyns}, which are illustrated in Figure~\ref{fig1}.  As already mentioned, the region $\Omega_D>0$ is future invariant. Similarly, since
\[
(Y'_{D})_{|_{Y_D=0}}=-\beta (1-H_D^2)^{3/2}<0,
\] 
the region $Y_D<0$ is future invariant. Hence if an orbit crosses the line $Y_D=0$ from the right to the left, 
it will remain on the left region for all future times. Furthermore, in the region $Y_D<0$ equation~\eqref{HDeq}   
implies that $H_D$ is monotone decreasing and that the subregion $H_D<0$ is future invariant. 
It follows easily that the $\omega$-limit of all orbits that cross the line  
$Y_D=0$  from the right to the left is the fixed point $\mathrm{F}_-$. 
Moreover if an orbit is entirely contained in the region $Y_D<0$, then the $\alpha$-limit of this orbit is 
the fixed point $\mathrm{F}_+$. 

Since the fixed point $\mathrm{S}_{(k)}$ is a saddle, it is the $\omega$-limit set of two interior orbits and 
the $\alpha$-limit set of two other interior orbits. These orbits divide the state-space in four regions, which 
we denote by \A, \B, \C, and \D; see Figure~\ref{fig1}. Each orbit in the region \A\, corresponds to solutions of the Einstein equations of type \A, as defined in Section~\ref{genclas} and similarly for the other regions.  The shadowed region in the pictures in Figure~\ref{fig1} identifies the region of accelerated expansion (if $H_D>0$) and contraction (if $H_D<0$), i.e., the region where $q<0$, which is 
\[
\Big(1-\frac{2}{3\gamma}\Big)(H^2_D+k(1-H_D^2))<Y_D<(H^2_D+k(1-H_D^2)),
\]
cf.~\eqref{qnew} and~\eqref{positiveOmega}. 
We continue by studying the qualitative behavior of the orbits in these regions separately for the cases $k=0$, $k=-1$ and $k=1$.

\subsubsection*{Case $k=0$}
\begin{figure}[Ht!]
\begin{center}
\psfrag{A}[cc][cc][1][0]{\A}
\psfrag{F+}[cc][cc][0.8][0]{$\mathbf{\mathrm{F}_+}$}
\psfrag{F-}[cc][cc][0.8][0]{$\mathbf{\mathrm{F}_-}$}
\psfrag{S}[cc][cc][0.8][0]{$\mathbf{\mathrm{S}_{(0)}}$}
\psfrag{dS+}[cc][cc][0.8][0]{$\mathbf{\mathrm{dS}_+}$}
\psfrag{dS-}[cc][cc][0.8][0]{$\mathbf{\mathrm{dS}_-}$}
\psfrag{M}[cb][cc][1][37]{de-Sitter}
\subfigure[Region A ($k=0$)]{\includegraphics[width=0.45\textwidth]{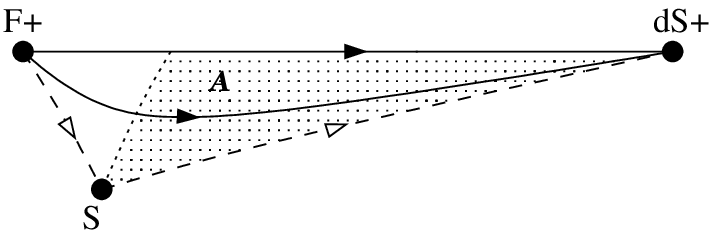}}\qquad
\psfrag{B}[cc][cc][1][0]{\B}
\subfigure[Region B ($k=0$)]{\includegraphics[width=0.45\textwidth]{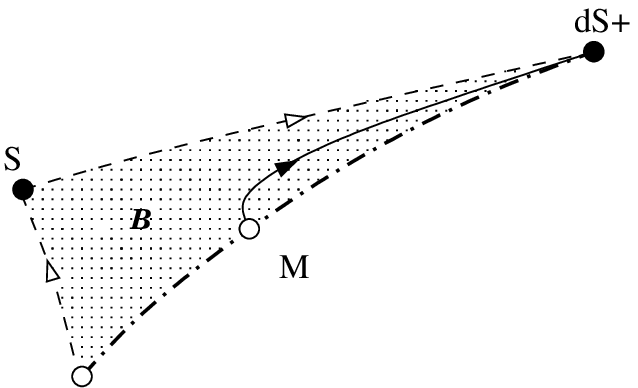}}\\
\psfrag{C}[cc][cc][1][0]{\C}
\psfrag{m1}[rl][tl][0.7][0]{\!Minkowski}
\psfrag{c1}[cc][cc][0.8][0]{$\mathbf{c_1}$}
\psfrag{c2}[cc][cc][0.8][0]{$\mathbf{c_2}$}
\psfrag{M1}[cc][cc][0.8][48]{de-Sitter}
\psfrag{M2}[rr][bb][0.8][-43]{de-Sitter}
\subfigure[Region C ($k=0$)]{\includegraphics[width=0.4\textwidth]{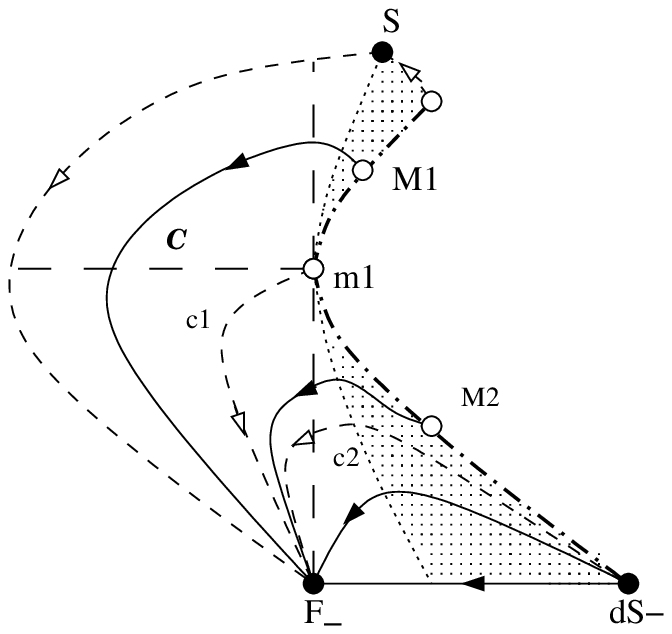}\label{figCflat}}\qquad
\psfrag{D}[cc][cc][1][0]{\D}
\subfigure[Region D ($k=0$)]{\includegraphics[width=0.4\textwidth]{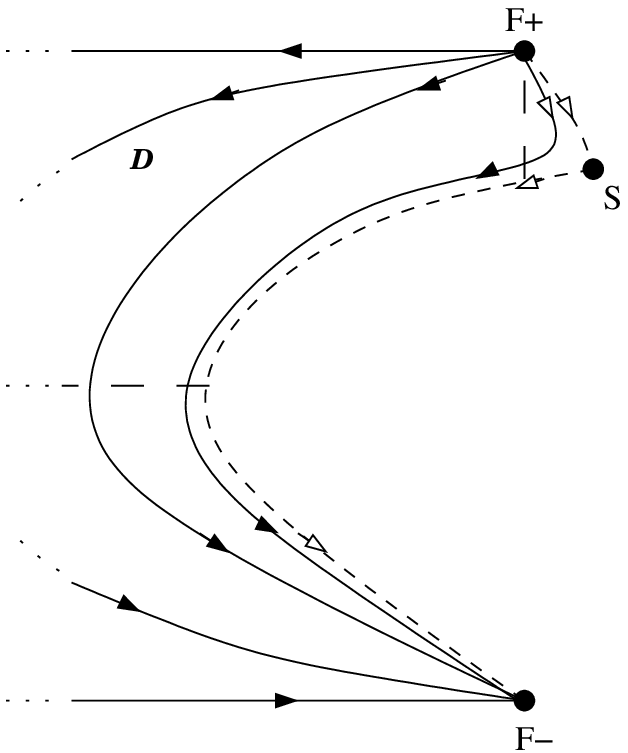}}
\end{center}
\caption{Phase portrait in the various regions of state-space ($k=0$)}\label{fig1many}
\end{figure}

The qualitative behavior of orbits in the various regions of the state-space in the case $k=0$ is depicted in Figure~\ref{fig1many} and can be described as follows.

The $\alpha$- and $\omega$-limit set of the orbits in the region \A~are the  fixed points $\mathrm{F}_+$ and $\mathrm{dS}_+$, respectively. Hence the solutions of the system~\eqref{system} corresponding to these orbits have a singularity in the past, toward which they behave like the expanding Friedman solution, while in the future they are singularity free and asymptotically de-Sitter. Moreover $H_D>0$ in the region \A~and all orbits enter the shadowed region before converging to $\mathrm{dS}_+$. Therefore the corresponding solutions of the Einstein equations are forever expanding and, after some finite time, the expansion becomes accelerated. 

All orbits in the region \B~converge to the fixed point $\mathrm{dS}_+$, hence the corresponding solutions of the Einstein equations are singularity free and asymptotically de-Sitter in the future time direction. In the past direction, orbits are forced to intersect the vacuum line $\Omega_D=0$. In terms of solutions to the system~\eqref{system} this means that, letting $t=0$ be the time of intersection, the scale factor can be continued to~\eqref{flatdesitter} (with $H_0>0$) for $t\leq 0$. The resulting cosmological model is then given by the de-Sitter spacetime for $t\leq 0$. Furthermore, since $H_D>0$ and the region \B~is completely shadowed, solutions corresponding to orbits in this region undergo accelerated expansion {\it for all times}.   

Orbits in the region \C~converge to the fixed point $\mathrm{F}_-$ in the future, which means that the corresponding solutions of the Einstein equations possess a singularity in the future toward which they behave like the contracting Friedman solution. Toward the past, orbits in the region \C~intersect the vacuum line $\Omega_D=0$. It is worth to distinguish the behavior of the corresponding solutions of the system~\eqref{system} into three different types. First we point out the existence of two special orbits in the region \C, which we call $c_1$ and $c_2$. The orbit $c_1$ passes through the point $(Y_D,H_D)=(0,0)$ (which corresponds to Minkowski), while the orbit $c_2$ originates from $\mathrm{dS}_+$ in the direction tangent to $\Omega_D=0$ at that point. The orbits $c_1,c_2$ divide the region \C~into three invariant regions, see Figure~\ref{figCflat}. Solutions corresponding to orbits below $c_2$ are singularity free and asymptotically de-Sitter in the past. Moreover they are forever contracting, 
because $H_D<0$. The orbits above $c_2$ intersect the vacuum line in the interior of the state-space 
(i.e., at finite time), and thus the corresponding solutions of the Einstein equations can be continued 
by de-Sitter prior to this time (i.e., by~\eqref{flatdesitter} with $H_0<0$). 
The difference between an orbit on the left of $c_1$ and an orbit on the right of $c_1$ is that the former corresponds to a solution which is initially expanding and then recollapsing (because the sign of $H_D$ changes from positive to negative along the orbit), while the latter corresponds to a forever contracting solution (since $H_D<0$ along the orbit.)      

Finally, the $\alpha$-limit set of orbits in the region \D~is the point $\mathrm{F}_+$, while the $\omega$-limit set is $\mathrm{F}_-$. This means that the corresponding solutions of the system~\eqref{system} become singular at finite time in both time directions. Moreover, since the sign of $H_D$ changes from positive to negative along all orbits, the solutions of the Einstein equations associated to the orbits in the region \D~are initially expanding, until they reach a stage of maximum extension, and then they recollapse into the future singularity.

\begin{figure}[Ht!]
\begin{center}
\psfrag{A}[cc][cc][1][0]{\A}
\psfrag{F+}[cc][cc][0.8][0]{$\mathbf{\mathrm{F}_+}$}
\psfrag{F-}[cc][cc][0.8][0]{$\mathbf{\mathrm{F}_-}$}
\psfrag{S}[cc][cc][0.8][0]{$\mathbf{\mathrm{S}_{(-1)}}$}
\psfrag{dS+}[cc][cc][0.8][0]{$\mathbf{\mathrm{dS}_+}$}
\psfrag{dS-}[cc][cc][0.8][0]{$\mathbf{\mathrm{dS}_-}$}
\psfrag{M}[cb][cc][1][37]{de-Sitter}
\psfrag{M2}[cc][cc][0.8][-20]{anti-de-Sitter}
\psfrag{M1}[cc][cc][0.8][20]{\ anti-de-Sitter}
\psfrag{M5}[rr][ll][0.8][-20]{de-Sitter}
\psfrag{M4}[cc][ll][0.8][20]{de-Sitter}
\psfrag{M3}[cc][cc][0.8][0]{Milne}
\psfrag{c1}[cc][cc][0.8][0]{$\mathbf{c_1}$}
\psfrag{c2}[bc][cc][0.8][0]{$\mathbf{c_2}$}
\psfrag{c3}[cc][cc][0.8][0]{$\mathbf{c_3}$}
\psfrag{c4}[cc][cc][0.8][0]{$\mathbf{c_4}$}
\psfrag{S-}[rr][bb][0.8][0]{\quad$\mathbf{\mathrm{S}_{(\!-\!1)}}$}
\subfigure[Region A ($k=-1$)]{\includegraphics[width=0.45\textwidth]{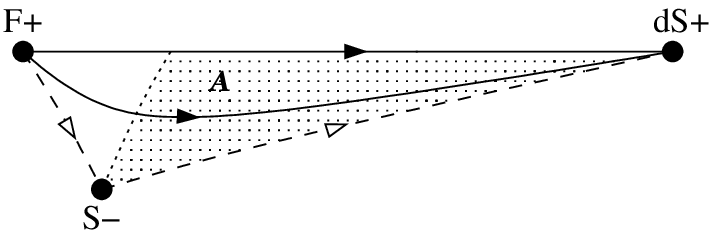}}\qquad
\psfrag{B}[cc][cc][1][0]{\B}
\subfigure[Region B ($k=-1$)]{\includegraphics[width=0.45\textwidth]{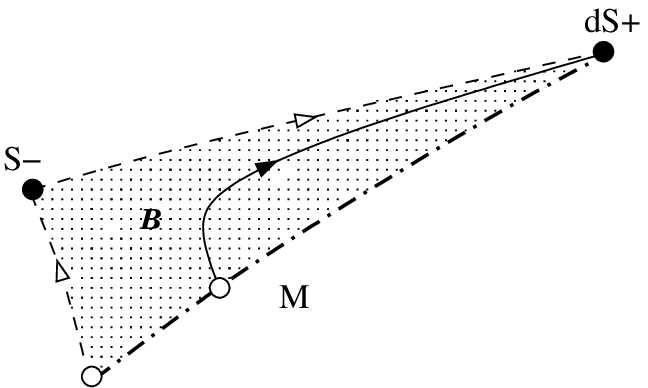}}\\
\psfrag{C}[ll][cc][1][0]{\!\C}
\subfigure[Region C ($k=-1$)]{\includegraphics[width=0.5\textwidth]{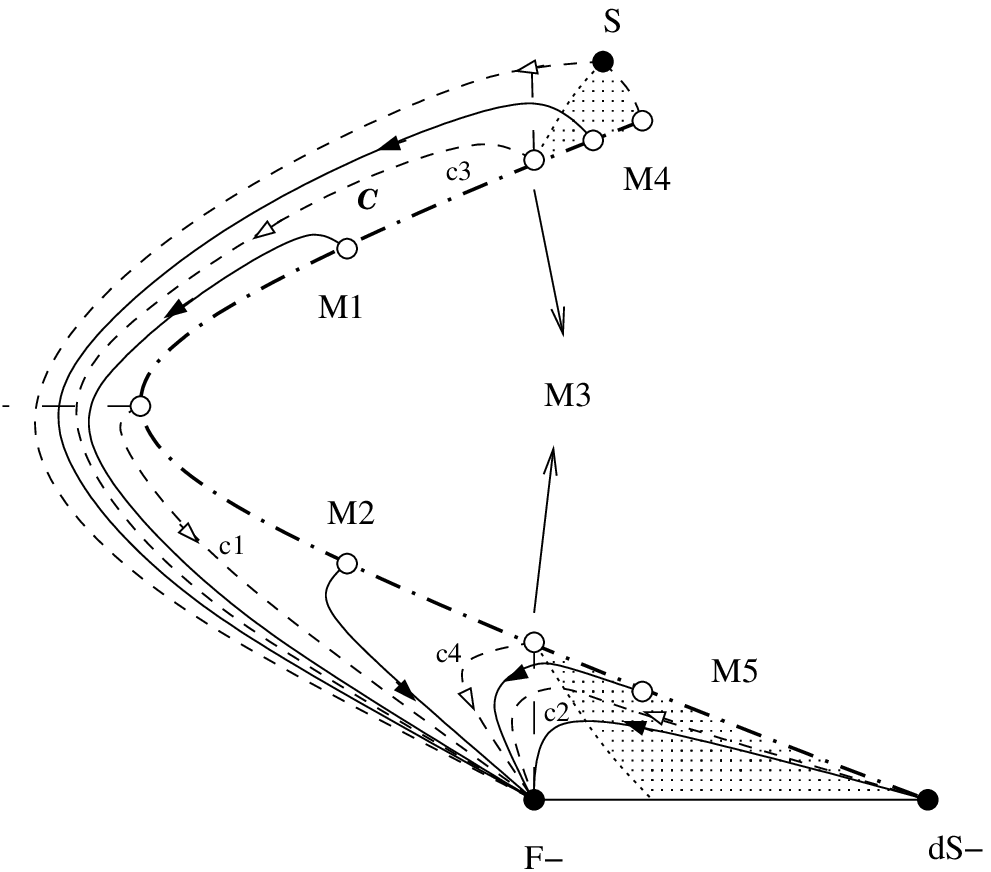}\label{figCopen}}\qquad
\psfrag{D}[cc][cc][1][0]{\D}
\subfigure[Region D ($k=-1$)]{\includegraphics[width=0.4\textwidth]{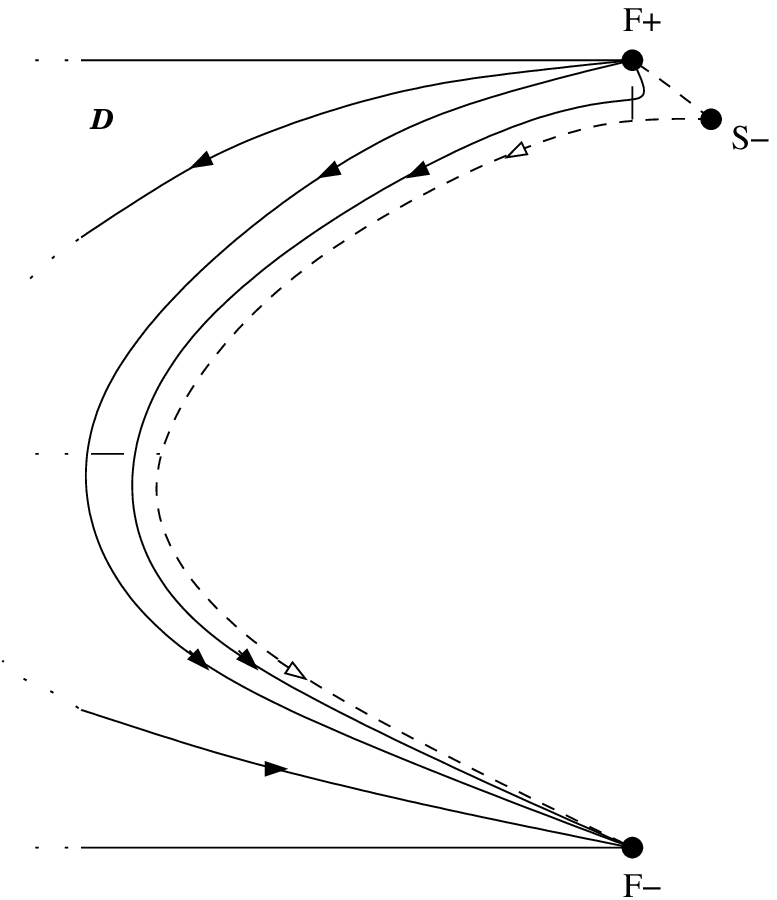}}
\end{center}
\caption{Phase portrait in the various regions of state-space ($k=-1$)}\label{fig2many}
\end{figure}

\begin{figure}[Ht!]
\centering
\psfrag{A}[cc][cc][1][0]{\A}
\psfrag{B}[cc][cc][1][0]{\B}
\psfrag{C}[cc][cc][1][0]{\C}
\psfrag{D}[cc][cc][1][0]{\D}
\psfrag{c1}[cc][cc][0.8][0]{$\mathbf{c_1}$}
\psfrag{c2}[bc][cc][0.8][0]{$\mathbf{c_2}$}
\psfrag{c5}[cc][cc][0.8][0]{$\mathbf{c_5}$}
\psfrag{c6}[cc][cc][0.8][0]{$\mathbf{c_6}$}
\psfrag{M1}[cc][cc][0.8][90]{de-Sitter}
\psfrag{M}[cc][cc][0.8][90]{de-Sitter}
\psfrag{Ff}[cc][bb][0.8][0]{$\mathbf{\mathrm{F}_-}$}
\psfrag{F-}[cc][cc][0.8][0]{$\mathbf{\mathrm{F}_-}$}
\psfrag{F+}[cc][cc][0.8][0]{$\mathbf{\mathrm{F}_+}$}
\psfrag{dS+}[cc][cc][0.8][0]{$\mathbf{\mathrm{dS}_+}$}
\psfrag{dS-}[cc][bb][0.8][0]{$\mathbf{\mathrm{dS}_-}$}
\psfrag{S}[rr][bc][0.8][0]{$\mathbf{\mathrm{S}_{(+1)}}$}
\subfigure[Region A ($k=1$)]{\includegraphics[width=0.45\textwidth]{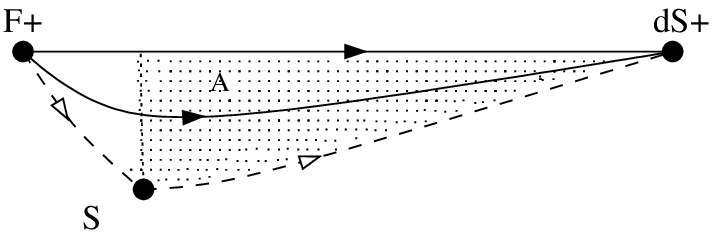} \label{fig3A} }    \qquad  
\subfigure[Region B ($k=1$)]{\includegraphics[width=0.25\textwidth]{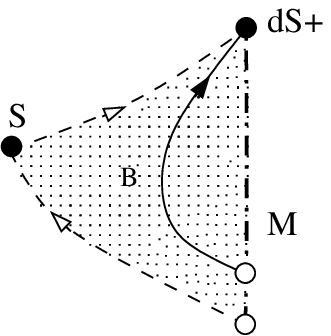} \label{fig3B} }    \bigskip   \\
\subfigure[Region C ($k=1$)]{ 
     \includegraphics[width=0.6\textwidth]{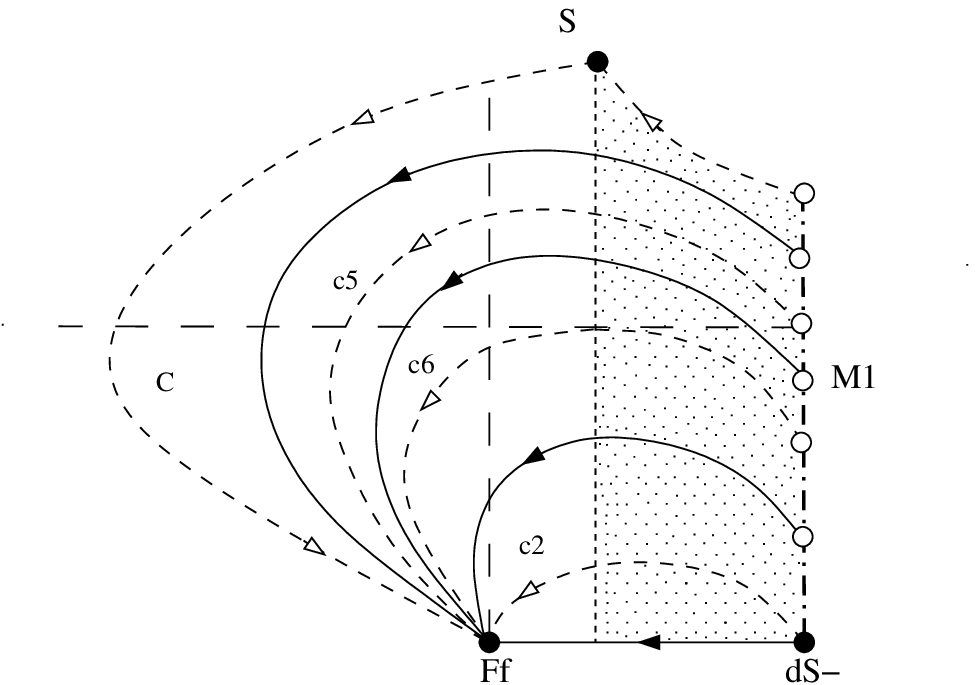}\label{fig3C} }   \
\subfigure[Region D ($k=1$)]{\includegraphics[width=0.35\textwidth]{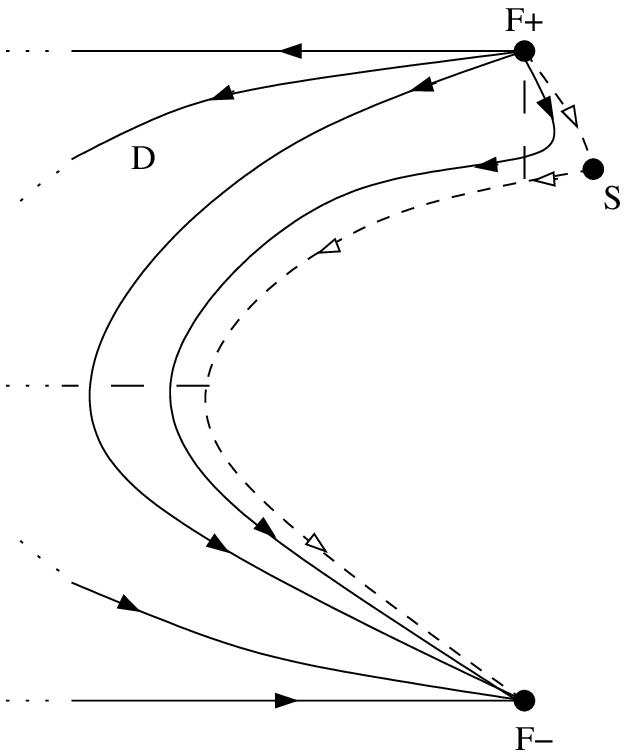} \label{fig3D} }
\caption{Phase portrait in the various regions of state-space ($k=1$)}\label{fig3many}
\end{figure}

\subsubsection*{Case $k=-1$}
The qualitative behavior of orbits for $k=-1$ is very similar to the case $k=0$, as it can be seen in 
Figure~\ref{fig2many}. We merely point out that in the region \C, depicted in Figure~\ref{figCopen}, the 
matching of the scale factor at $\Omega_D=0$ is either with de-Sitter (i.e.,~\eqref{opendeSitter}), or with 
Milne (i.e.,~\eqref{milne}), or with anti-de-Sitter (i.e.,~\eqref{antidesitter}). In particular, there exists 
two special orbits, which we call respectively $c_3$ and $c_4$, intersecting the vacuum line in the expanding 
and contracting Milne solution. Both these orbits converge to $\mathrm{F}_-$ in the future and divide the region 
\C\ into three parts. The solutions of the Einstein equation corresponding to orbits in the interior part 
delimited by $c_3$ and $c_4$ are matched to the anti-de-Sitter solution, while the the solutions corresponding 
to the two exterior regions above $c_2$ are matched with de-Sitter. 
 Below $c_2$, the solutions are asymptotic to de-Sitter in the past.

\subsubsection*{Case $k=1$}
The qualitative behavior of the orbits in regions \A, \B\ and \D\ for this case is similar to the previous cases $k=0$ and $k=-1$, 
as shown in Figure~\ref{fig3many}. 
As for region \C, depicted in Figure~\ref{fig3C},
the matching at  $\Omega_D=0$ of the orbits above the orbit $c_2$ 
is with de-Sitter (i.e.,~\eqref{opendeSitter}), while below $c_2$ 
the solutions are asymptotic to de-Sitter in the past. 
In this case, there are two special orbits, $c_5$ and $c_6$, intersecting the vacuum line (on the boundary of the phase space) 
in the expanding and contracting de-Sitter solutions, respectively. 
Both these orbits converge to $\mathrm{F}_-$ in the future and divide the region \C\ into three parts. 
The interior part delimited by $c_5$ and $c_6$ consists of orbits that correspond to solutions which are
initially contracting, then expanding and finally recollapsing (because the sign of $H_D$ changes from negative to positive and
then from positive to negative along the orbits).
The orbits between $c_2$ and $c_6$ are forever contracting (since $H_D<0$ along the orbits).
The orbits above $c_5$ are initially expanding and then collapsing (because the sign of $H_D$ changes from positive to negative along the orbits).


\section{Summary}
\label{sec:summ}

In this work we considered a cosmological model based on the Einstein equations and the assumption that the matter content of the universe is described by a fluid undergoing diffusion in a cosmological scalar field. The scalar field can be identified with the dark energy field. We take spacetime to have a Robertson-Walker line element, so that the model studied here is 
spatially homogeneous and isotropic. The matter field variables are solutions of a non linear system of ordinary 
differential equations. We were able to obtain all solutions in which the scale factor is linear on time. 
In order to understand the dynamical properties of general solutions, we rewrite the system in terms of 
normalized (dimensionless) dynamical variables. We have shown that typical solutions of the Einstein equations  can be classified 
according to their past and future asymptotic behavior into four classes, which we called \A, \B, \C, \D. 
 Solutions of type \A\,  describe a forever expanding universe 
having a Big Bang singularity in the past and which is asymptotic to de-Sitter toward the future; type \B\, 
solutions describe a singularity-free cosmological model which at some finite time in the past can be matched to 
de-Sitter. Solutions of this type are forever expanding with acceleration, and since the scalar field is monotonically 
decreasing, the future asymptotic de-Sitter state has a smaller value of the cosmological constant 
than the past de-Sitter matching solution. Solutions of type \C\, and \D\, have a Big Crunch sigularity in the 
future, due to the  scalar field becoming negative and thus mimicking a negative cosmological constant, but while 
solutions of type \C\, have a Big Bang singularity in the past, those of type \D\, are singularity free and 
can be matched at finite time (or approach asymptotically) either de-Sitter or Anti-de-Sitter depending on the 
value of the curvature parameter $k$.


\section*{Acknowledgements}
AA is supported by CAMGSD, Instituto Superior T{\'e}cnico through the project EXCL/MAT-GEO/0222/2012 from 
the ``Funda\c c\~ao para a Ci\^encia e a Tecnologia'' (FCT) as well as by the FCT grant SFRH/BPD/85194/2012. Furthermore, AA  thanks the
Department of Mathematics at Chalmers University, Sweden, for
the kind hospitality. 
MPMR and AJS are supported  by
the Research Centre of Mathematics of the University of Minho
through the FCT Projects PEst-C/MAT/UI0013/2011 and PEstOE/MAT/UI0013/2014. 

\




\begin{thebibliography}{99}

\bibitem{AU2015} 
A.~Alho, and C. Uggla: Global dynamics and inflationary center manifold and slow-roll approximants. J. Math. Phys. {\bf 56}, 012502 (2015) 


\bibitem{AHU2015} 
A.~Alho, J. Hell, and C. Uggla: Global dynamics and asymptotics for monomial scalar field potentials and perfect fluids. Preprint 2015

\bibitem{simone1}
S. Calogero: A kinetic theory of diffusion in general relativity with cosmological scalar field. JCAP 11/2011, 016 (2011)

\bibitem{simone2}
S. Calogero: Cosmological models with fluid matter undergoing velocity diffusion. J. Geom. Phys.
{\bf 62}, 2208--2213 (2012)

\bibitem{CV} S.~Calogero, H.~Vetten: Cosmology with matter diffusion. JCAP 11/2013, 025 (2013)

\bibitem{CH} S.~Calogero, J.~M.~Heinzle: Bianchi Cosmologies with Anisotropic Matter: Locally Rotationally Symmetric Models. Physica D {\bf 240}, 636--669 (2011)

\bibitem{Coley} A.~A.~Coley: {\it Dynamical systems and cosmology}. Kluwer Academic Publishers (2003)

\bibitem{F98a} S.~Foster: Scalar Field Cosmologies and the Initial Space-Time Singularity. Class. Quant. Grav.
{\bf 15}, 3485-3504 (1998)

\bibitem{F98b} S.~Foster: Scalar Field Cosmological Models With Hard Potential Walls. Available at arXiv:gr-qc/9806113



\bibitem{Gib} G.~W.~Gibbons: Anti-de-Sitter and Its Uses. In {\it Mathematical and Quantum Aspects of Relativity and Cosmology}. Lecture Notes in Physics {\bf 537}, 102--142 (2000)

\bibitem{GE99}
M. Goliath, G.F.R. Ellis: Homogeneous cosmologies with a cosmological constant. Phys. Rev. D
{\bf 60}, 023502 (1999)

\bibitem{HU}J.~M.~Heinzle, C.~Uggla: Dynamics of the spatially homogeneous Bianchi type I Einstein-Vlasov equations. Class.~Quant.~Grav. {\bf 23}, 3463--3490 (2006)

\bibitem{Kim} Y.~Kim, C.~Y.~Oh, N.~Park: Classical Geometry of de-Sitter Spacetime: An Introduction Review. J.~Korean Phys.~Soc. {\bf 42}, 573--592 (2003)

\bibitem{Perko}
L. Perko:
\textit{Differential Equations and Dynamical Systems (Third Edition)} Springer- Verlag, New York (2000)







\bibitem{ri} M.~Risken:
\emph{The Fokker-Planck Equation: Methods of Solutions and Applications}. Springer Series in Synergetics {\bf 18}, Springer-Verlag, Berlin (1996) 





\bibitem{shogin} D.~Shogin, S.~Hervik:  Evolution of a Simple Inhomogeneous Anisotropic Cosmological Model with Diffusion. JCAP 10/2013, 005 (2013) 
\bibitem{shogin2} D.~Shogin, S.~Hervik: Dynamics of tilted Bianchi models of types III, IV, V in presence of diffusion. Preprint arXiv:1402.2785 
\bibitem{shogin3} D.~Shogin, S.~Hervik: The late-time behaviour of tilted Bianchi type VIII universes in presence of diffusion. Class. Quantum Grav. {\bf 31} 135006 (2014)

\bibitem{U13} 
C.~Uggla: Global cosmological dynamics for the scalar field representation of the modified Chaplygin gas. 
Phys. Rev. D {\bf 88}, 064040 (2013)


\bibitem{WE}
J. Wainwright and G.F.R. Ellis (Ed.): 
\textit{Dynamical systems in cosmology}
Cambridge University Press (1997)

\end{thebibliography}
\end{document}